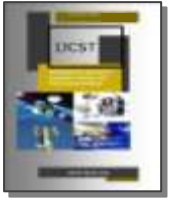

# MEEP: Multihop Energy Efficient Protocol For Heterogeneous Wireless Sensor Network

Surender Kumar[1], M. Prateek[2], N.J. Ahuja[3] and B. Bhushan[4]

[1-3]Centre of Information Technology, University of Petroleum & Energy Studies, Dehradun, India
[4]G.N.Khalsa College, Yamunanagar, India
[1]skmalan@gmail.com, [2]mprateek@ddn.upes.ac.in, [3]neelu@ddn.upes.ac.in, [4]bharat_dhiman@sify.com

*Abstract*— Energy conservation of sensor nodes for increasing the network life is the most crucial design goal while developing efficient routing protocol for wireless sensor networks. Recent technological advances help in the development of wide variety of sensor nodes. Heterogeneity takes the advantage of different types of sensor nodes and improves the energy efficiency and network life. Generally sensors are deployed randomly and densely in a sensing region so short distance multihop communication reduces the long distance transmission in the sensor network. In this research paper MEEP (multihop energy efficient protocol for heterogeneous sensor network) is proposed. The proposed protocol combines the idea of clustering and multihop communication. Heterogeneity is created in the network by using some nodes of high energy. Low energy nodes use a residual energy based scheme to become a cluster head. High energy nodes act as the relay nodes for low energy cluster head when they are not performing the duty of a cluster head to save their energy further. Protocol also suggests a sleep state for nodes in the cluster formation process for saving energy and increasing the life of sensor network. Simulation result shows that the proposed scheme is better than other two level heterogeneous sensor network protocol like SEP in energy efficiency and network life.

*Index Terms*— Cluster, Energy-Efficiency Multihop, Network Lifetime, Residual Energy, Wireless Sensor Network

## I. INTRODUCTION

RECENT advances of microelectronic and wireless communication make WSNs a promising technology for wide variety of civil and military applications. Wireless sensor network consists of a large number of tiny sensors which are deployed densely and randomly without any preplanning in a region [1], [2], [3], [4], [5]. Sensor nodes are limited in storage space, computing power and energy. Battery of the nodes cannot be changed or recharged due to their large number and harsh environment deployment. Thus energy saving is an important design goal for wireless sensor network.

Clusters based routing helps in solving the sensors energy constraints by reducing the cost of data transmission and aggregation of sensed data before transmitting data to base station [4]. Clustering divides the sensing region in a hierarchical way to create clusters and communication between nodes are controlled by cluster heads. Cluster head is either randomly selected from nodes by using a probability scheme or a centralized control algorithm is used for this purpose so that the energy load can be evenly distributed in sensor network [5]. Thus clustering using a two layer approach in the sensor network where first layer is used for selecting the cluster head and the second layer for routing [4]. With the introduction of heterogeneity in sensor network energy can be further saved without sacrificing performance. There are three common types of resource heterogeneity in sensor nodes: computational heterogeneity, link heterogeneity and energy heterogeneity [5]. Nodes in a wireless sensor network are deployed densely therefore multihop communication is desirable for achieving the energy efficiency. A node joins a cluster which is near to it depending upon join message signal strength however in some cases this choice is not energy efficient. To get rid of this problem a sleep state has been introduced during the cluster formation process which further increases the network life and energy efficiency. In this research paper a multihop and clustering scheme are combined together and suggests a new multihop energy efficient protocol to improve sensor network life and energy saving. Rest of this paper is organized in to the sections as follows. Section 2 describes the related work of cluster based protocols, section 3 explains the wireless sensor network model and section 4 describes the radio energy model which has been selected for this paper, section 5 describes the proposed MEEP protocol, section 6 explores the simulation results and finally the paper is concluded in section 7.

## II. RELEATED WORK

Clustering schemes are of two types. In homogeneous network all the nodes are of same type and if some of the nodes are of different type then the network is known as heterogeneous. In recent years a number of cluster based routing protocols have been proposed for sensor network. LEACH (Low energy adaptive clustering hierarchy) [7] is one of the important cluster based protocol which uses an adaptive approach for the cluster head election. Every node generates a random number between 0 and 1 and when this number is less than a particular threshold value T (n) (as shown in Equation (1) then the node becomes a cluster head for the current round

After becoming a cluster head node advertise about its status in the network and non cluster head nodes join a nearby cluster head. Cluster head further creates a TDMA/CSMA schedule for its members to avoid intracluster and intercluster collision. In LEACH there is no guarantee about number and placement of cluster heads in a round. LEACH-C (LEACH Centralized) [7] is a protocol which uses base station control algorithm for the cluster head elections and disperse the cluster heads in the entire sensing region. Initially each sensor node sends the information about their location and energy by using a GPS to base station. On the basis of this information base station finds the average energy of the network and the nodes having energy below this cannot become a cluster head during a round. LEACH-C is more efficient than LEACH and delivers 40 % more data per unit energy.

$$T(n) = \begin{cases} \frac{p}{1-p \times \left(r \bmod \frac{1}{p}\right)} & if\ n\ \in G \\ 0 & otherwise \end{cases} \quad (1)$$

Here G denotes the set of nodes that are not selected as a cluster head in last $\frac{1}{p}$ rounds and r is the current round.

One of the disadvantages of the LEACH is that the cluster heads rotations do not take into account the remaining energy of sensor nodes. A node may not have sufficient energy to complete a round and may be selected as a CH. In [8] a new approach for cluster head selection is proposed. When the remaining energy of a node is larger than 50% of initial energy then for cluster head election LEACH algorithm is used as in Equation (1). Otherwise a new scheme which considers the remaining energy of each node is applied for cluster head selection according to Equation (2).

$$T(n) = \begin{cases} \frac{p}{1-p \times (r \bmod \frac{1}{p})} \times \left(2p \times \frac{E_{residual}}{E_{init}}\right) & if\ n\ \in G \\ 0 & otherwise \end{cases} \quad (2)$$

Here p is the percentage of nodes that can become cluster head, $E_{residual}$ is the remaining energy and $E_{init}$ is the initial energy of a node and G is the set of nodes that have not become cluster head in the last $\frac{1}{p}$ rounds and r is the current round.

PEGASIS [9] is a greedy chain based algorithm for data gathering in wireless sensor network. In PEGASIS each node forms a chain structure through which the data is forwarded to base station. PEGASIS achieves energy efficiency by transmitting data to only one of its neighbor node where the collected data is fused and further forwarded to next one hop neighbor. There is no rapid depletion of the energy of the base station nearer nodes because all the nodes are doing the data fusion at its place.

HEED [10] is another popular distributed energy efficient clustering algorithm which probabilistically elects cluster heads based on their residual energy and nodes join cluster heads which have the minimum communication. One of the key features of HEED is that it exploits the multiple transmission power levels of sensor nodes. TEEN [11] is a cluster based protocol for time critical applications and uses two threshold values hard and soft for the election of cluster heads. Hard and soft threshold determines the minimum value and changes that are of interest in s sensing region. As a result number of transmissions to base station is reduced considerably.

SEP [12] is a cluster based protocol for two level heterogeneous network. In SEP there are two types of nodes: normal and advanced. Advanced nodes have more energy than the normal nodes and it is the source of heterogeneity in the network. Weighted probabilities of normal and advanced nodes are used to determine the thresholds for the election of cluster head in a round.

DEEC [13] is a distributed energy efficient clustering protocol for heterogeneous network in which nodes become the cluster head based on the residual and the average energy of the network. Energy expenditure of nodes is controlled by means of an adaptive approach and for this average energy is used as the reference energy.

EEHC [14] purposes an energy efficient scheme for heterogeneous wireless sensor network. EEHC increases the life of the network by 10 % as compared to LEACH in the presence of powerful node setting. EDEEC [15] is a clustering algorithm with three types of nodes and uses residual and average energy of the network for the selection of cluster heads in a round.

EECDA [16] is another cluster based protocol for three level heterogeneous networks. Some percentage of the nodes of network has more energy than normal nodes which are known as advanced nodes. Further in advanced nodes some fractions of the nodes have even more energy than the normal nodes which are known as super nodes. Novel cluster head election and a path of maximum sum of energy residues for data transmission is used in EECDA for increasing network lifetime and stable region.

DBCP [17] is an energy efficient clustered protocol for heterogeneous wireless sensor network which selects the cluster heads according to initial energy and average distance

### III. HETEROGENEOUS NETWORK MODEL

This section describes the wireless sensor network model of the proposed protocol. Model contains n sensor nodes which are randomly deployed in a 100 × 100 square meters region as shown in Figure 1. Various assumptions of the network model and sensor nodes are as follows.

- Nodes are deployed randomly and uniformly in the sensing region.
- Base station and nodes become stationary after deployment and base station is located in the middle of sensing region.
- Nodes are location unaware i.e. they do not have any information about their location.
- Nodes continuously sense the region and they always have the data to send the base station.
- Battery of the nodes cannot be changed or recharged due to harsh environment deployment.

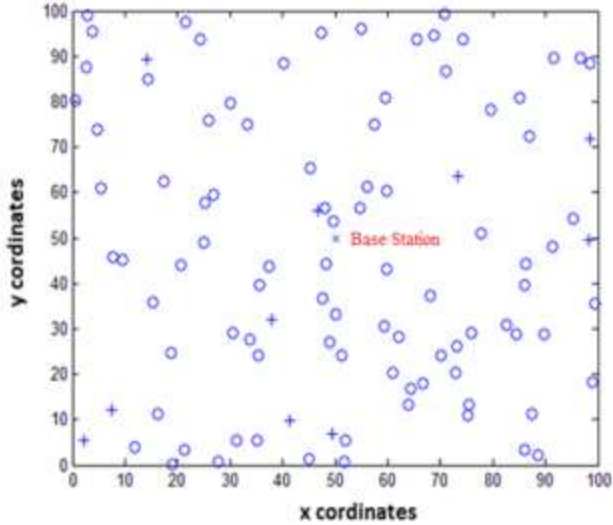

**Figure 1: Sensing Region**
**(o - Normal, + - Advanced Node, x – Base Station)**

Sensing region has two types of sensor nodes i.e. advanced and normal nodes. Let initial energy of normal nodes is $E_0$ and m be the fraction of advanced nodes having initial energy $E_0 \times (1+\alpha)$, where α means that advanced node have α times more energy than normal node. Heterogeneous network total initial energy is given by

$$E_{Total} = N \cdot (1-m)E_0 + N \cdot mE_0(1+\alpha) = N \cdot E_0(1+\alpha m) \quad (3)$$

In the analysis radio energy model similar to [7]. Free space ($d^2$ power loss) and the multipath fading ($d^4$ power loss) channel model both are used depending upon the distance between the transmitter and receiver. When distance is less than specific a threshold value then free space model are used otherwise multipath loss model is used. The amount of energy required to transmit *L* bit packet over a distance, *d is* given by Equation (4).

$$E_{Tx}(L,d) = \begin{cases} L \times E_{elec} + L \times \varepsilon_{fs} \times d^2 & if\ (d < d_0) \\ L \times E_{elec} + L \times \varepsilon_{mp} \times d^4 & if\ (d \geq d_0) \end{cases} \quad (4)$$

$E_{elec}$ is the electricity dissipated to run the transmitter or receiver. The parameters $\varepsilon_{mp}$ and $\varepsilon_{fs}$ is the amount of energy dissipated per bit in the radio frequency amplifier according to the distance $d_0$ which is given by the Equation (5).

$$d_0 = \sqrt{\frac{\varepsilon_{fs}}{\varepsilon_{mp}}} \quad (5)$$

For receiving an *L* bit message the energy expends by radio is given by

$$E_{Rx}(L) = L \times E_{elec} \quad (6)$$

## IV. MEEP

MEEP (Multihop Energy Efficient Protocol for Heterogeneous Wireless Sensor Networks) is a clustering protocol for two level heterogeneous sensor networks. MEEP consists of n sensors and base station is located in the middle f the sensing region. The distance of any node to its cluster head or sink is ≤ $d_0$. The energy dissipated in the cluster head during a single round is given by Equation (7).

$$E_{CH} = \left(\frac{n}{k} - 1\right) \times L \cdot E_{elec} + \frac{n}{k} L \cdot E_{DA} + L \cdot E_{elec} + L \cdot \varepsilon_{fs} \cdot d_{toBS}^2 \quad (7)$$

Where L is the no of bits of the data message, $d_{toBS}$ is the average distance between base station and cluster head and $E_{DA}$ is the energy required for data aggregation in a round. Since cluster members transmit data to its cluster head therefore energy dissipated in a non cluster head follows the free space path and it is given by Equation (8)

$$E_{NCH} = L \times (E_{elec} + \varepsilon_{fs} \times d_{toCH}^2) \quad (8)$$

Where $d_{toCH}$ is the average distance between node and cluster head and the energy depleted in a round is given by

$$E_{Cluster} = E_{CH} + \left(\frac{n}{k} - 1\right) E_{NCH} \approx E_{CH} + \left(\frac{n}{k}\right) E_{NCH} \quad (9)$$

The total energy dissipated in the network is given by Equations (10)

$$E_T = L \times (2nE_{elec} + nE_{DA} + k\varepsilon_{fs}d_{toBS}^2 + n\varepsilon_{fs}d_{toCH}^2) \quad (10)$$

The optimal number of clusters can be found by finding the derivative of $E_{Total}$ with respect to k and equating it to zero.

$$k_{opt} = \frac{\sqrt{n}}{\sqrt{2\pi}} \sqrt{\frac{\varepsilon_{fs}}{\varepsilon_{mp}}} \frac{M}{d_{toBS}^2} \quad (11)$$

The average distance from cluster head to the sink can be calculated in the following way [12]

$$d_{toBS} = 0.765 \frac{M}{2} \quad (12)$$

Node's optimal probability to become cluster head in round is given by Equation (13).

$$p_{opt} = \frac{k_{opt}}{n} \quad (13)$$

Epoch of the network increases in proportion of the energy increment. Heterogeneous nodes increases the system energy by $\alpha \cdot m$ times hence for optimizing the stable region, new epoch must become equal to $\frac{1}{p_{opt}} \cdot (1 + \alpha \cdot m)$ [10]. Let $p_{nrm}, p_{adv}$ denotes the weighted election probability for normal and advanced nodes respectively. The average number of cluster heads per round per epoch is equal to $(n \times (1 + \alpha \cdot m) \cdot p_{nrm})$. Normal and advanced nodes weighted probability can be calculated by using the equation 14 and 15 respectively.

$$p_{nrm} = \frac{p_{opt}}{1+\alpha \cdot m} \quad (14)$$

$$p_{adv} = \frac{p_{opt}}{1+\alpha \cdot m} \times (1 + \alpha) \quad (15)$$

In a homogenous network, to guarantee that there is average $n \times p_{opt}$ cluster-heads in every round, each node $s_i$ becomes a cluster-head once every $n_i = \frac{1}{p_{opt}}$ rounds. When the network operates for some time then the nodes cannot have the same residual energy. Thus energy is not well distributed if the rotating epoch $n_i$ is equal for all nodes. Hence low-energy nodes will die quickly than high-energy nodes of the network. To get rid of this problem MEEP protocol chooses different $n_i$ for normal nodes on the basis of residual energy $E_i(r)$ of node $s_i$ in a round r.

Let $p_i = \frac{1}{n_i}$ be average probability to become a cluster-head during $n_i$ rounds. When nodes have the same energy at each epoch, choosing the average probability $p_i$ to be $p_{opt}$ can ensure that there are $n \times p_{opt}$ cluster-heads in each round and all nodes die approximately at the same time. If nodes have different energy, $p_i$ of the nodes with more energy should be larger than $p_{opt}$. Let $E_i(t)$ denotes the initial energy and $E_i(r)$ represents residual energy of a normal node $s_i$ at round r, using $E_i(t)$ to be the reference energy, for normal nodes we have

$$p_{opt} = p_{opt} \frac{E_i(r)}{E_i(t)} \quad (16)$$

$p_{opt}$ is the reference value of average probability and determines the threshold $T(s_i)$ of node $s_i$. In two-level heterogeneous networks, MEEP replace the reference value $p_{opt}$ with the weighted probabilities given in Equation (13) for normal nodes however for advanced nodes the weighted probabilities remain the same as for SEP. Thus $p_i$ is changed into.

$$p_i = \begin{cases} \frac{p_{opt} \times E_i(r)}{(1+\alpha \cdot m) \times E_i(t)} & if\ s_i\ is\ the\ normal\ node \\ \\ \frac{p_{opt} \cdot (1+\alpha)}{(1+\alpha \cdot m)} & if\ si\ the\ advanced\ node \end{cases} \quad (17)$$

The probability threshold $T(s_i)$ which node $s_i$ uses to determine whether it can become a cluster-head in a round can be calculated by using the Equation 18.

$$T(s_i) = \begin{cases} \frac{p_i}{1-p_i\left(r\ mod\ \frac{1}{p_i}\right)} & if\ s_i\ \varepsilon\ G' \\ \\ 0 & otherwise \end{cases} \quad (18)$$

Depending upon whether $s_i$ is a normal or advanced node, $G'$ represents the set of either normal or advanced nodes that are not elected as cluster heads within the last $\frac{1}{p_{nrm}}$ or $\frac{1}{p_{adv}}$ rounds. Further to reduce the energy consumption of wireless sensor network MEEP introduces three tier architectures concept for normal sensor nodes. In a round if a normal sensor node becomes a cluster head then after collecting the data from its members it aggregates the data and instead of sending the data directly to sink it will try to find out an advanced node such that

- That is not a cluster head in this particular round.
- Distance between normal cluster head and advanced node is less than the distance between normal cluster head and base station.

If the normal cluster head find such advanced node that is not a cluster head in this round r and also its distance is less than distance between normal cluster head and base station then normal cluster head instead of sending the data directly to the base station it sends its data to this advanced node which further send it to the base station. If normal cluster head does not find any such advanced node who fulfils the above mentioned two conditions then it will send the aggregated data of its members directly to the base station itself. Thus by introducing a gateway concept or three tier architectures for normal cluster head, MEEP has reduced the energy consumed in transmission to prolong the network lifetime and stability period. Moreover, when the cluster-heads are selected, each node joins to the closest (considering the transmission power) cluster-head. However in some cases this is not the optimal choice because, if exists a sensor node in the base station direction whose distance from the base station is less than all the nearby cluster-head distance (Figure 2).

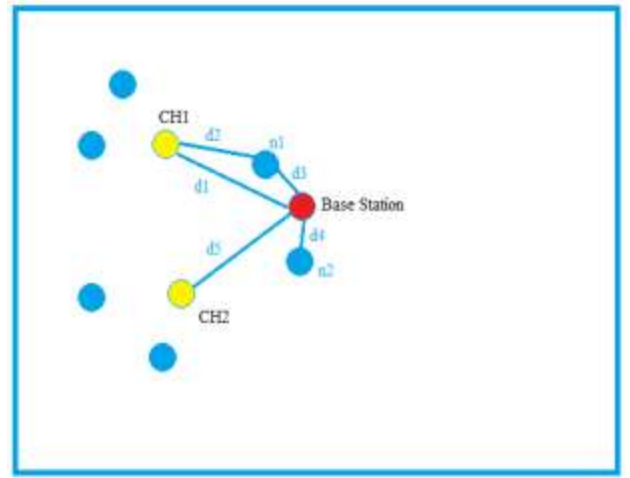

**Figure 2: Cluster Formation Process**

Let us consider the figure 2, where the node n1 has to transmit L-bits message to the base station. The closest cluster-head to n1 is CH1. And, if the node belongs to this cluster, it spends energy (19).

$$E1 = L \cdot E_{elec} + L \cdot \epsilon_{d2} \cdot d_2^x \qquad (19)$$

Where
$$\begin{cases} x = 2, \epsilon_{d2} = 10 \text{ pj / bit / m}^2, & if\ d_2 < d_0 \\ x = 4, \epsilon_{d2} = 0.0013 \text{ pj / bit / m}^4, & if\ d_2 \geq d_0 \end{cases}$$

But if the node n1 chooses to transfer data to the base station directly, this energy will be:

$$E2 = L \cdot E_{elec} + L \cdot \epsilon_{d3} \cdot d_3^y \qquad (20)$$

Where
$$\begin{cases} y = 2, \epsilon_{d3} = 10 \text{ pj / bit / m}^2, & if\ d_3 < d_0 \\ y = 4, \epsilon_{d3} = 0.0013 \text{ pj / bit / m}^4, & if\ d_3 \geq d_0 \end{cases}$$

Here positive coefficients $x\ and\ y$ represent the energy dissipation radio model used. Clearly $E2 < E1$ but in this case lot of uncompressed data is collected at the base station. To get rid of this problem we have introduced a sleep state in MEEP in the following manner. When $E1 > E2$ clearly it is not an optimal choice for saving energy and transmit data. Thus in this case instead of sending the data to the cluster head *CH1*, sensor node *n1* enters into a sleep state and waits for the next round in which it either itself becomes a cluster head or find a nearby cluster such that $E1 < E2$. Sensor node *n1* remains in the sleep state for the maximum 4 rounds, if in these next 4 rounds it either becomes a cluster head or finds a nearby cluster such that $E1 < E2$ then it wakes up and performs their respective duty either of a cluster head or the member of a cluster head. If sensor node *n1* is neither able to become the members of a cluster such that $E1 < E2$ nor become a cluster head in the next 4 rounds, then sensor node wakes up and transmit the data directly to the base station.

## V.  SIMULATION AND RESULTS

The performance of MEEP is compared with SEP. For simulation 100 x 100 square meters region with 100 sensor nodes is used in which base station is located in middle of the region as shown in Figure 1 The normal nodes are represented by the symbol (o), advanced nodes (Gateway) with (+) and the base Station by (x). The radio parameters used for the simulations are given in TABLE 1.The following metrics is used to evaluate the performance of the proposed algorithm.

(i) **Stability Period:** This is the time interval between network start until the death of the first node

(ii) **Network lifetime:** This is the time interval between network start until the death of the last node

(iii) **Number of cluster heads per round:** This will reflect the number of cluster heads formed in each round.

(iv) **Number of Alive nodes per round:** This will measure the total number of live nodes per round.

(v) **Throughput:** This will measure the total number of packets which are sent to base station.

**Table 1: Radio Parameters of MEEP**

| Parameter | Value |
|---|---|
| $E_{elec}$ | 5 nJ/bit |
| $\varepsilon_{fs}$ | 10 pJ/bit/m$^2$ |
| $\varepsilon_{mp}$ | 0.0013 pJ/bit/m$^4$ |
| $E_0$ | 0.5 J |
| $E_{DA}$ | 5 nJ/bit/message |
| Message Size | 4000 bits |
| $p_{opt}$ | 0.1 |
| $d_0$ | 70m |

The proposed protocol is tested by introducing various parameters of heterogeneity in the system and the following cases are considered for heterogeneity

Case 1: m = 0.1, a = 5
Case 2: m = 0.2, a = 3

**Case 1: m = 0.1, a =5**

In this case 10 advanced nodes are deployed with 5 times more energy than normal nodes and there are 90 normal nodes. Figure 3 shows that network lifetime of MEEP is more than SEP as last node dies later in it. First node dies earlier in SEP than MEEP therefore stability period of SEP is less than MEEP. Figure 4 shows that no of alive nodes per round are more in MEEP than SEP. Figure 5 show that throughput i.e. total number of messages send to base station is more in MEEP. Figure 6 plots the total remaining energy in joules per round and remaining energy per round of MEEP is also more than SEP. Thus MEEP is more robust than SEP.

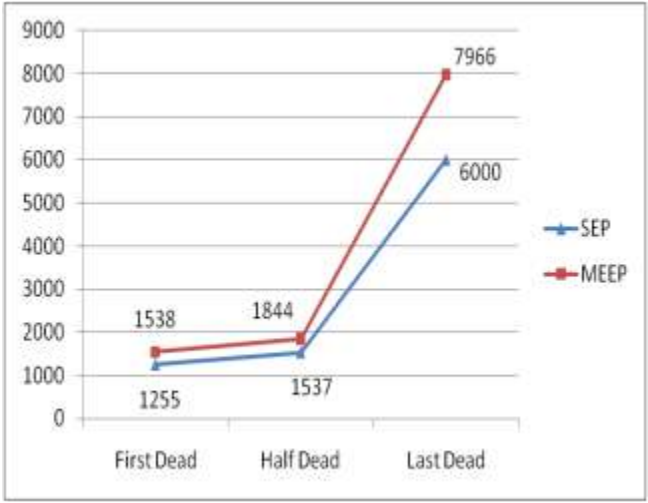

**Figure 3: Rounds for 1st, half and Last node death in MEEP & SEP**

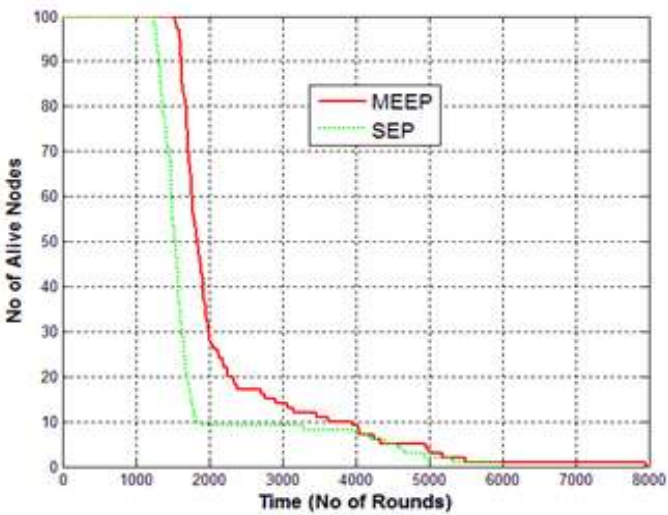

**Figure 4: Comparison of Alive nodes in MEEP & SEP**

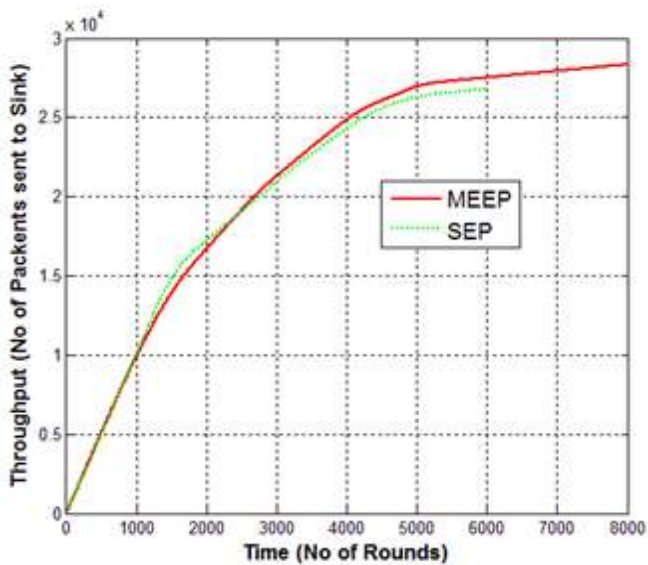

**Figure 5: Comparison of Throughput in MEEP & SEP**

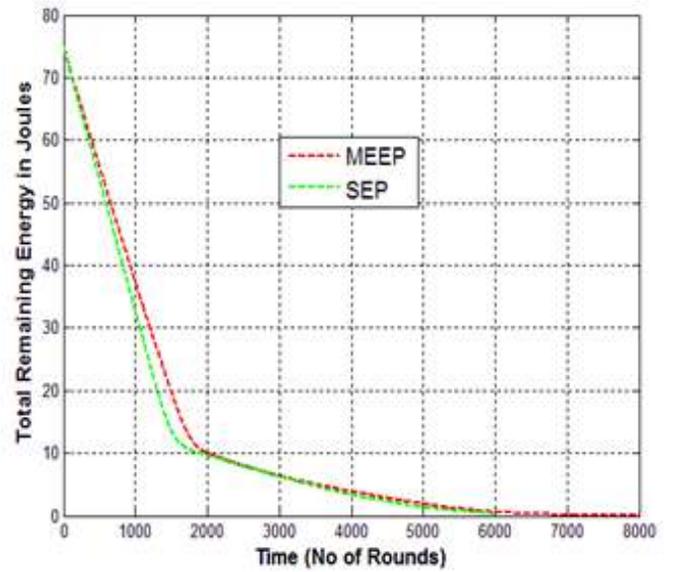

**Figure 6: Comparison of Total remaining energy in MEEP & SEP**

**Case 2: m = 0.2, a = 3**

In this case 20 advanced nodes are deployed with 3 times more energy than normal nodes and there are 80 normal nodes. Figure 7 shows that network lifetime of MEEP is more than SEP as last node dies earlier in SEP. Stability period of MEEP is also more than SEP because first node dies later in it. Figure 8 show that no of alive nodes per round are more in MEEP than SEP. Figure 9 shows that throughput i.e. total number of packets send to base station is also more in MEEP. Figure 10 plots the total remaining energy in joules per round and this is also more in MEEP as compared to SEP.

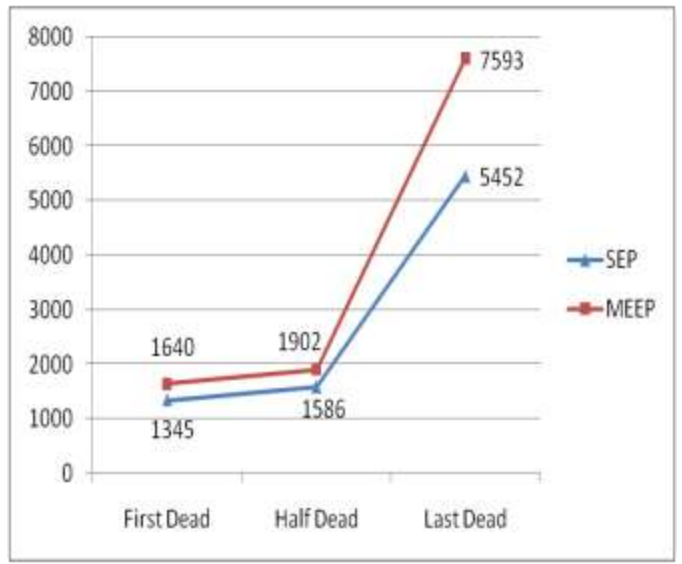

**Figure 7: Rounds for 1st, half and Last node death in MEEP & SEP**

## VI. Conclusion

MEEP is a multihop energy efficient clustering algorithm for heterogeneous sensor network. The proposed protocol is an extension of SEP and it takes the full advantage of heterogeneity. It improves the network lifetime, stable region and throughput of the network. For taking the full advantage of heterogeneity MEEP introduces a multihop architecture for normal cluster heads. Advanced nodes further takes over the data transmission load of normal cluster head to save network energy, prolong the stable region, network lifetime and throughput of the system. When MEEP is compared with SEP it increases the sensor network stable region, throughput, and network lifetime approximately by   22 %, 35% and 39 % respectively.

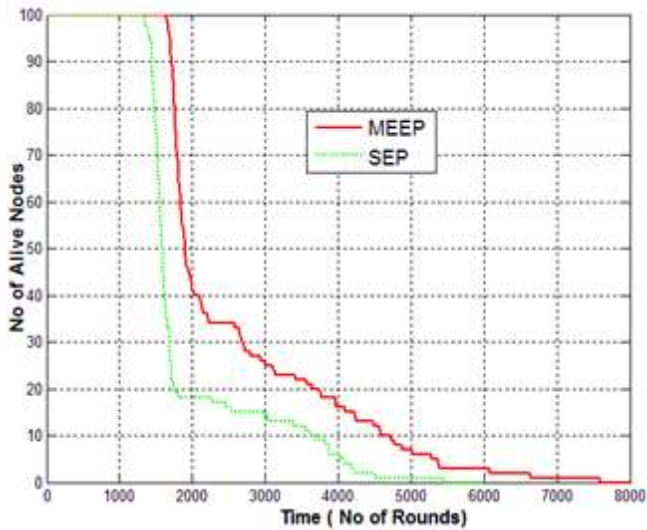

**Figure 8: Comparison of Alive nodes in MEEP & SEP**

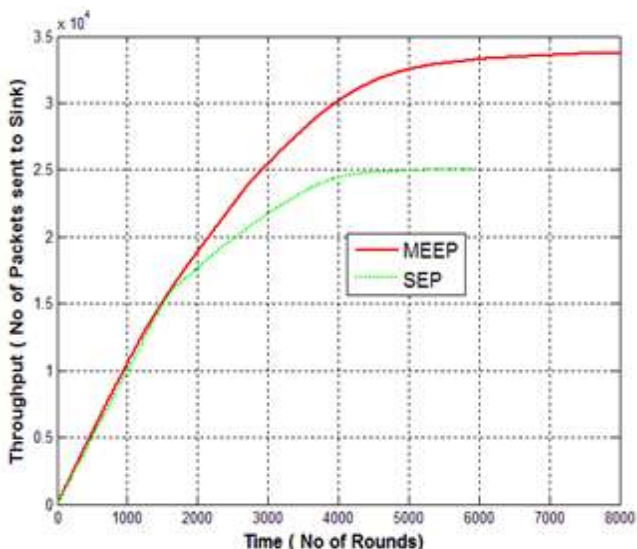

**Figure 9: Comparison of Throughput in MEEP & SEP**

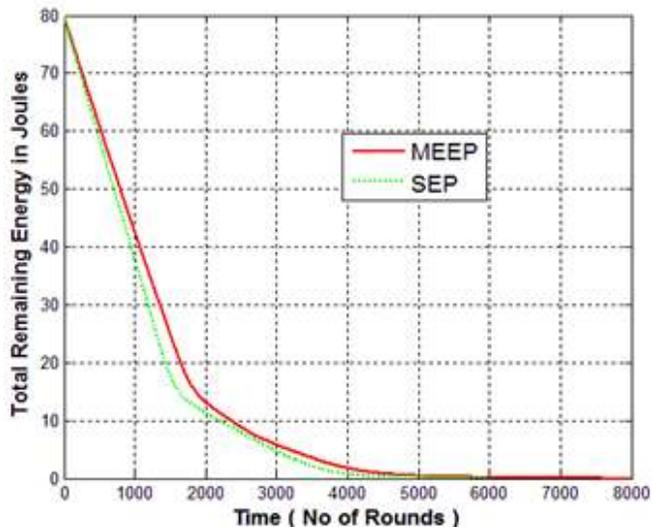

**Figure 10: Comparison of Total remaining energy in MEEP & SEP**

## REFERENCES

[1] J. Yick, B. Mukherjee, and D. Ghosal, "Wireless sensor network survey", Computer Networks, Vol. 52, Issue 12, pp. 2292-2330, August 2008

[2] IF. Akyildiz, W. Su, Y. Sankarasubramaniam, and E. Cayirci, "Wireless sensor networks: a survey," Computer Networks; vol. 38, issue 4, pp. 393–422, March 2002.

[3] Kemal Akkaya and Mohamed Younis, "A Survey on Routing Protocols for Wireless Sensor Networks," Ad Hoc Networks, Vol. 3, No. 3, pp. 325-349, May 2005

[4] J. Al-Karaki, and A. Kamal, "Routing Techniques in Wireless Sensor Networks: A Survey," IEEE Communications Magazine, Vol. 11, no. 6, Dec. 2004, pp. 6-28

[5] M. Yarvis, N. Kushalnagar, H. Singh "Exploiting heterogeneity in sensor networks," IEEE INFOCOM, 2005

[6] Jun Zheng and Abbas Jamalipour, "Wireless Sensor Networks: A Networking Perspective," a book published by A John & Sons, Inc, and IEEE, 2009

[7] W. Heinzelman, A. Chandrakasan, and H. Balakrishnan, "An application specific protocol architecture for wireless microsensor networks,"  IEEE Transactions on Wireless Communications , vol. 1, no. 4 , pp. 660 – 670, Oct. 2002

[8] Jang, Ki Young, Kyung Tae Kim, Hee Yong Youn, "An Energy Efficient Routing Scheme for Wireless Sensor Networks", Computational Science and its Application 2007. ICCSA 2007." In International Conference on Volume. Issue, pp. 26-29. 2007.

[9] S. Lindsey and C. Raghavendra, "PEGASIS: Power-Efficient Gathering in Sensor Information Systems," IEEE Aerospace Conf. Proc., 2002, vol. 3, 9–16, pp. 1125–30.

[10] Younis, Ossama and Sonia Fahmy. "HEED: a hybrid, energy-efficient, distributed clustering approach for ad hoc sensor networks." Mobile Computing, IEEE Transactions on 3, no. 4 (2004),  pp. 366-379

[11] A. Manjeshwar and D. P. Agrawal , " TEEN: A routing protocol for enhanced efficiency in wireless sensor networks ," Proceedings IPDPS ' 01 , San Francisco, CA, Apr.2001, pp 2009 – 2015.

[12] G. Smaragdakis, I. Matta, A. Bestavros, "SEP: A Stable Election Protocol For Clustered Heterogeneous Wireless Sensor Networks," in proceedings of 2nd International Workshop on Sensor and Actor Network Protocols and Applications (SANPA'04), Boston, MA, 2004, pp. 660-670.

[13] Qing, Li, Qingxin Zhu, and Mingwen Wang. "Design of a distributed energy-efficient clustering algorithm for

heterogeneous wireless sensor networks." Computer Communications 29, no. 12, pp. 2230-2237, 2006.

[14] Kumar Dilip, Trilok C. Aseri, and R. B. Patel, "EEHC: Energy efficient heterogeneous clustered scheme for wireless sensor networks," Computer Communications 32, no. 4, pp. 662-667, 2009.

[15] Saini, Parul, and Ajay K. Sharma. "E-DEEC-enhanced distributed energy efficient clustering scheme for heterogeneous WSN." In Parallel Distributed and Grid Computing (PDGC), 2010 1st International Conference on, IEEE, 2010, pp. 205-210.

[16] D.Kumar, T.C. Aseri, R.B. Patel, "EECDA: energy efficient clustering and data aggregation protocol for heterogeneous wireless sensor networks," International Journal of Computers Communications & Control, 6(1), pp. 113-124, 2011.

[17] S.Kumar, M. Prateek, B. Bhushan,"Distance based cluster protocol for heterogeneous wireless sensor network," IJCA, vol. 76, No. 9, pp. 42-47, August 2013.

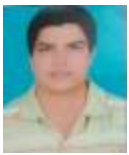

**Surender Kumar** received his Master of Computer Applications degree from Department of Computer Science & Applications, Kurukshetra University, India in 2001. He received his M.Phil in Computer Science degree from Madurai Kamraj University, India in 2007 and now is a PhD student at College of Engineering Studies in University of Petroleum and Energy Studies, India. His research interests include sensor networks, computer network and network security.

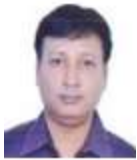

**Dr. Manish Prateek** received his PhD from Lalit Narayan Mithila University, India. He did his research work in the area of Manufacturing and Robotics from the Memorial University of Newfoundland, Canada in 2001. He is currently working as Professor in the College of Engineering Studies, University of Petroleum & Energy Studies, India. He published a number of research papers in various prestigious conference proceedings and Journals like IEEE, IDMME, CANCAM (IEEE) and NECEC. His research interests include robotics, embedded systems,data structure network..

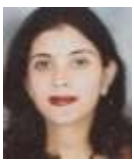

**Dr. Neelu Jyothi Ahuja** received her PhD from University of Petroleum &Energy Studies, India . She is currently working as Assistant Professor in the College of Engineering Studies, University of Petroleum & Energy Studies. She has more than 14 years experience and published many research papers in journals and conferences at National and International level. Her research interests include expert systems, artificial intelligence, object oriented development and programming language.

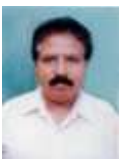

**Dr. Bharat Bhushan** received his PhD from Kurukshetra University, India. He is currently working as Head, Department of Computer Science & Applications, Guru Nanak Khalsa College, Yamunanagar, India. He has more than 29 years of experience and published many research papers in journals and conferences at International and National level. His research interest includes software engineering, digital electronics, networking and simulation experiments.